\documentclass[preprint,aps,12pt,preprintnumbers,eqsecnum,nofootinbib]{revtex4}
\usepackage{graphicx}
\usepackage{subfigure}
\usepackage{epstopdf} 
\usepackage{braket}
\usepackage{amssymb,amsmath,amsfonts,comment,latexsym,graphicx,epstopdf,bbold,float}
\usepackage{color}
\usepackage{fancyvrb}
\usepackage{chngcntr}
\usepackage{etoolbox}
\usepackage{bbm,epsfig}
\usepackage{array}
\usepackage{hepunits}

\newcommand{\be}{\begin{equation}}
\newcommand{\ee}{\end{equation}}
\newcommand{\ba}{\begin{eqnarray}}
\newcommand{\ea}{\end{eqnarray}}

\newcommand{\grts}{\raise.3ex\hbox{$>$\kern-.75em\lower1ex\hbox{$\sim$}}}
\newcommand{\lets}{\raise.3ex\hbox{$<$\kern-.75em\lower1ex\hbox{$\sim$}}}

\def\braket#1{\mathinner{\langle{#1}\rangle}}

{\catcode`\|=\active\gdef\Braket#1{\left<\mathcode`\|"8000\let|\bravert {#1}\right>}}

\def\bravert{\egroup\,\vrule\,\bgroup}

\usepackage{subfigure}

\usepackage{color}
\usepackage{amssymb,amsmath,amsfonts}
\usepackage{epstopdf} 
\usepackage{braket}

\begin{document}
%
%
\title{\vspace*{0.5in} 
$T'$ Models with High Quality Flaxions
\vskip 0.1in}
\author{Christopher D. Carone}\email[]{cdcaro@wm.edu}
\author{Marco Merchand}\email[]{mamerchandmedi@email.wm.edu}

\affiliation{High Energy Theory Group, Department of Physics,
William \& Mary, Williamsburg, VA 23187-8795}

%
\date{April 12, 2020}
\begin{abstract}
 The ``gauged" Peccei-Quinn (PQ) mechanism of Fukuda, Ibe, Suzuki and Yanagida is implemented in the flavorful axion model of Carone and 
 Merchand.   This model of flavor is similar to other successful ones based on the double tetrahedral group, but the flavor symmetry includes a 
 global U(1) factor that leads to the presence of a flavorful axion.   Here we gauge that U(1) symmetry and introduce a heavy sector that includes
(1) the fermions necessary to cancel anomalies and (2) a second scalar flavon field that spontaneously breaks the U(1) symmetry.  The full 
theory has an accidental U(1)$\times$U(1)$'$ global  symmetry, anomalous with respect to QCD; U(1)$_\text{PQ}$ emerges as a linear 
combination.   The gauged flavor symmetry restricts the possible PQ symmetry-breaking higher-dimension operators so that sufficient axion 
quality is preserved.    We provide a model  of the quark sector, as a proof of principle, and then a model which incorporates the standard model 
charged leptons as well.  In both cases, the charge assignments that lead to acceptable axion quality also lead to a multiplicity of some of the 
heavy sector states; we check that the Landau pole for hypercharge remains above the cut off of the effective theory.   We consider relevant 
phenomenological constraints on these models including those on the predicted axion couplings.
\end{abstract}

\maketitle

\section{Introduction} \label{sec:intro}
The absence of a CP-violating term quadratic in the gluon field-strength tensor, {\em i.e.} one proportional to 
$G_{\mu\nu} \widetilde{G}^{\mu\nu}$, remains one of the puzzles of the standard model.   The Peccei-Quinn 
mechanism~\cite{Peccei:1977hh,Peccei:1977ur} posits the existence of a spontaneously broken U(1) global symmetry, 
one that is anomalous with respect to QCD;  the goldstone boson of this symmetry, the axion, couples to 
$G_{\mu\nu} \widetilde{G}^{\mu\nu}$ so that this term vanishes when the axion sits at the minimum of its non-perturbatively 
generated potential.  Aside from providing a dynamical mechanism for solving the strong-CP problem, the axion is also a 
plausible dark matter candidate~\cite{Kolb:1990vq}.   Interest in axions has been heightened by the absence of compelling 
evidence for TeV-scale beyond-the-standard-model physics at the LHC, as well as the null results from dark matter 
experiments that search directly for weakly interacting massive particles at or around the electroweak scale.

Following earlier work~\cite{early}, one possibility that has reemerged recently is that the anomalous global symmetry of the Peccei-Quinn 
mechanism may play  a role in explaining the flavor structure of the standard model~\cite{Bjorkeroth:2018dzu,Calibbi:2016hwq,Arias-Aragon:2017eww,Linster:2018avp,Ema:2016ops,Carone:2019lfc,Ahn:2018cau}.  In the simplest models, a U(1)$_F$ flavor symmetry, 
spontaneously broken by a ``flavon" field $\varphi$, provides an origin for both the axion field and Yukawa coupling hierarchies~\cite{Ema:2016ops}.   The Yukawa couplings (aside from that of the top quark) arise via higher-dimension operators of the form
\begin{equation}
\frac{1}{M_F^p}\overline{Q}_L H \varphi^p d_R + \mbox{ h.c.} \,\, ,
\end{equation}
where we have used a charge $-1/3$ quark as an example, and where $p$ depends on the U(1)$_F$ charge assignments of the 
fields. By judicious choices of these assignments, Yukawa matrix entries can arise in a hierarchical pattern, as determined by the
powers $p$ that control the various entries, as well as the ratio $\langle  \varphi \rangle / M_F$ which is taken to be a small parameter. 
The axion $a$ can be identified using the nonlinear representation
\begin{equation}
\varphi = \frac{1}{\sqrt{2}}(\sigma+f) e^{i \, a / f}  \,\,\, ,
\end{equation}
where $\langle \varphi \rangle = f/\sqrt{2}$, and where $\sigma$ is a heavy field, with mass of ${\cal O}(f)$, that we will ignore.  Non-linear 
redefinitions of the fermion fields can remove $a$ from the Yukawa couplings and shift it to the fermion kinetic terms, where it will  
appear as a field that is derivatively coupled to the U(1)$_F$ Noether current.  Since the axion has couplings that are 
flavor-dependent, it has been called a ``flavorful axion,"~\cite{Bjorkeroth:2018dzu}
``axiflavon,"~\cite{Calibbi:2016hwq,Arias-Aragon:2017eww,Linster:2018avp} or
 ``flaxion,"~\cite{Ema:2016ops}  depending on the tastes of the authors.  We will refer to this type of axion 
as a flaxion in the present work.

Interesting models of flavor that involve non-Abelian groups may include discrete and/or continuous Abelian factors.   
Non-Abelian groups often lead to more predictive models than purely Abelian ones, since the fermions can be embedded in 
representations with dimension greater than $1$.  (By contrast, there is significantly more freedom when one can assign a U(1) 
charge to each fermion field independently.)   Nevertheless, Abelian factors are often necessary in these models, as is the case 
in a number of elegant models based on the double tetrahedral group $T'$.  For example, the supersymmetric models of 
Ref.~\cite{Aranda:1999kc,Aranda:2000tm} based on $T' \times Z_3$ require the $Z_3$ factor so that a subgroup exists that rotates 
the standard model fermion fields of the first generation (which reside within a $T'$ doublet) by a phase.  The subsequent breaking 
of this subgroup at a  lower energy scale accounts for the smallness of the Yukawa couplings of the first generation fermions relative 
to the other generations.   Differences between the up- and down-quark Yukawa matrices require additional symmetries, for example 
promoting $Z_3$ to $Z_3 \times Z_2$ in one of the models of Ref.~\cite{Aranda:2000tm}; the non-supersymmetric $T'$ 
flavor models studied in Ref.~\cite{Carone:2016xsi}, on the other hand, utilized $T' \times Z_3 \times Z_3$.    In Ref.~\cite{Carone:2019lfc}, similar nonsupersymmetric models were studied in which the second $Z_3$ factor was promoted to U(1), endowing the model with a 
flaxion to address the strong CP problem.   (For supersymmetric flaxion models based on $T'$ symmetry, see Ref.~\cite{Ahn:2018cau}.)  However, no origin was provided for this U(1) symmetry, which was assumed to arise as an 
artifact of some unspecified theory in the ultraviolet.  One nontrivial feature of such a completion is that it would have to 
solve the axion quality problem, {\em i.e.}, the problem that the Peccei-Quinn mechanism is easily rendered ineffective by 
quantum gravitational effects~\cite{Barr:1992qq}.  It is believed that quantum gravitational effects generally break all global 
symmetries~\cite{Kallosh:1995hi}, in this case through operators that would re-introduce the strong CP problem by 
triggering a non-zero value of the $\overline{\theta}$ parameter.  We will review this more explicitly later.  Possible mechanisms 
of producing ``high quality" axions have been 
proposed~\cite{Fukuda:2017ylt,Kim:1984pt,Choi:1985cb,Redi:2016esr,Lillard:2017cwx,Lillard:2018fdt,Duerr:2017amf}, but little 
discussion exists (as far as we are aware) in the context of flavored axion models.   

In this work, we consider flaxion quality in the context of the $T' \times Z_3 \times $U(1)$_F$ flaxion model discussed in 
Ref.~\cite{Carone:2019lfc}. We go beyond that work by building variant models with heavy sectors that make explicit the physics 
that protects the flaxion quality.  In particular, we promote the U(1)$_F$ factor of the original model to a gauge symmetry, and 
show that the extended model has an approximate U(1)$\times$U(1)$'$ global symmetry in which the U(1)$_F$ gauge group is 
embedded.   One of the two goldstone bosons that arise from the spontaneous breaking of the global symmetry becomes 
the longitudinal component of the U(1)$_F$ gauge boson, while the other remains as a flaxion.   The gauged flavor symmetry 
restricts the possible higher-dimension operators that can break the U(1)$\times$U(1)$'$ global symmetry so that flaxion 
quality is sufficiently preserved.   Thus, we present models that show how to successfully implement the ``gauged Peccei-Quinn" 
approach proposed by Fukuda, {\em et al.}~\cite{Fukuda:2017ylt} to the $T' \times Z_3 \times$U(1)$_F$  flaxion model of 
Ref.~\cite{Carone:2019lfc}.   This places the results of that work on sounder theoretical footing.

Our paper is organized as follows.  In Sec.~\ref{sec:background}, we give a brief summary of the Yukawa textures that emerge in the
$T' \times Z_3 \times $U(1)$_F$ model of Ref.~\cite{Carone:2019lfc}.  We will not need to review how the breaking of the 
$T' \times Z_3$ symmetry leads to most of the features of these matrices, since only the factors associated with the breaking of 
the U(1)$_F$ factor will be relevant  to our later discussion.  In Sec.~\ref{sec:quarks}, we present a model in which the 
$T' \times Z_3 \times$U(1)$_F$ symmetry is applied only to the quark sector; this model is consistent with a wide range of other 
possible flavor groups that might be relevant in the lepton sector. In Sec.~\ref{sec:leptons}, we consider a flaxion model in which the 
same flavor symmetry is relevant to both the quark and lepton sectors.  In Sec.~\ref{sec:conc}, we summarize our conclusions.

\section{Textures} \label{sec:background}
The models of interest are based on the flavor group $T^\prime \times Z_3 \times$U(1)$_F$.  The flavor symmetry breaking 
fields fall in $T'$ singlet and doublet representations; using the $T' \times Z_3$ notation of Ref.~\cite{Aranda:2000tm}, 
\begin{equation}
\phi \sim {\bf 2}^{0+} \,\,\, , \,\,\,\,\, A \sim {\bf 1}^{0-} \,\,\, ,  \mbox{ and } \,\,\, s \sim {\bf 1}^{00} \,\,\ .
\end{equation}
Details of $T'$ group theory, including an explanation of this notation and the Clebsch-Gordan matrices necessary for constructing 
invariant Lagrangian terms, can be found in Ref.~\cite{Aranda:2000tm}.  However, this will not be relevant to our subsequent 
discussion.  One only needs to know the Yukawa textures generated via the breaking of the flavor symmetry and the role played by the $s$ flavon, which is the only one that is charged under U(1)$_F$;  we choose our normalization so that this charge is $+1$.  More specifically, the symmetry breaking of the discrete factors is given by
\begin{equation}
T' \times Z_3 \stackrel{\epsilon}{\longrightarrow} Z_3^D \stackrel{\epsilon'}{\longrightarrow} \mbox{ nothing} \,\,\, ,
\end{equation}
where $Z_3^D$ refers to a diagonal subgroup of a $Z_3$ subgroup of $T'$ and the additional $Z_3$ factor~\cite{Aranda:1999kc,Aranda:2000tm}.  The dimensionless parameters $\epsilon$ and $\epsilon'$  are defined in terms of the symmetry-breaking vacuum expectation values (vevs) 
and the flavor scale $M_F$, the cut off of the effective theory:
\begin{equation}
\braket{\phi}/M_F \equiv \left[\begin{array}{c} \epsilon \\ 0 \end{array} \right]\, ,\,\,\,\,\, \braket{A} / M_F \equiv \epsilon' \, , \,\,\,\, \mbox{ and  }\,\,\,\, \braket{s}/M_F \equiv
 \rho \,\,\, .  \label{eq:vevpat}
\end{equation}
The additional dimensionless parameter $\rho$ is determined by the U(1)$_F$ breaking scale.   This leads to the leading-order
Yukawa textures
 \begin{equation}
Y_{U} \sim  \begin{pmatrix}
0      &     u_1  \epsilon'& 0 \\
-u_1  \epsilon'  &   u_2  \epsilon^2   & u_3 \epsilon \\ 
0 & u_4 \epsilon  & u_5
\end{pmatrix} , \label{structureU}
\end{equation}
\begin{equation}
Y_{D} \sim \begin{pmatrix}
0      &     d_1   \epsilon'  & 0 \\
-d_1  \epsilon'  &   d_2  \epsilon^2   & d_3 \epsilon \, \rho \\ 
0 & d_4 \epsilon  & d_5 \rho
\end{pmatrix},  \label{structureD}
\end{equation}  
\begin{equation}
Y_{E} \sim \begin{pmatrix}
0      &     l_1   \epsilon'  & 0 \\
-l_1  \epsilon'  &    l_2  \epsilon^2   & l_3  \epsilon  \\ 
0 & l_4  \epsilon\,  \rho & l_5  \rho
\end{pmatrix}, \label{structureE}  
\end{equation}
where the $u_i$, $d_i$ and $l_i$ are (in general complex) $\mathcal{O}(1)$ parameters.  These Yukawa matrices were 
shown to be phenomenologically viable in Ref.~\cite{Carone:2019lfc} via a global fit to the quark and lepton masses and 
the Cabibbo-Kobayashi-Maskawa (CKM) mixing angles;  in this fit, the values of the free parameters were consistent with the
expectations of naive dimensional analysis~\cite{NDA}.

\section{Quark Sector Model} \label{sec:quarks}

We focus in this section on a $T^\prime \times Z_3 \times$U(1)$_F$ model of quark flavor, corresponding to the quark sector 
of the model of Ref.~\cite{Carone:2019lfc}.   An extension to the lepton sector that assumes the same flavor group is presented 
in Sec.~\ref{sec:leptons}.   The quark-sector model presented in this section exemplifies our approach more directly, 
and is compatible with models of lepton flavor that may assume a different lepton flavor group structure.

The U(1)$_F$ in Ref.~\cite{Carone:2019lfc} was a global flavor symmetry whose spontaneous breaking at the flavor 
scale provided an origin for a flavored axion.   This breaking was accomplished by a single flavon field $s$, whose flavor 
charge was normalized to $+1$.  Of the quark fields, only the right-handed bottom quark carried a flavor charge, $-1$, so 
that Yukawa matrix entries that multiply $d_R^3$ acquire a suppression factor given by $\langle s \rangle/M_F \equiv \rho$, where $M_F$ 
was the flavor scale.   This factor, taken in addition to those related to the breaking of the $T^\prime$ symmetry, provides for the 
successful Yukawa textures that were summarized in the previous section.  Since the U(1)$_F$ symmetry is anomalous 
with respect to color, the flavored goldstone boson that emerges from spontaneous symmetry breaking serves as a 
viable flavored axion. 

To implement the ``gauged" Peccei-Quinn idea of Fukuda, {\em et al.}~\cite{Fukuda:2017ylt}, we introduce another flavon 
field $s'$, with U(1)$_F$ charge $-1/N$, with $N$ an integer to be determined later.  This field will couple to $N$ heavy colored 
states $D_R^j$ and $D_L^j$, for $j=1 \ldots N$.    We promote this symmetry to a gauged flavor symmetry.   We will see that 
at leading order in a $1/M_F$ expansion, the theory including the heavy sector fields has an enlarged global symmetry, 
U(1)$\times$ U(1)$^\prime$, corresponding to separate phase rotations on the $s$ and $s'$ fields.  Gauging the U(1)$_F$ 
flavor symmetry leaves the full theory with a residual U(1) global symmetry that is both anomalous and spontaneously broken, 
assuring the presence of a flavorful axion.  However, a consequence of the gauged flavor symmetry is that the set of operators that break the residual global symmetry explicitly occur only at very high order, so that the 
flavored axion evades the axion quality problem.  In this section we assume the simplest possibility, that the flavor scale is identified with the reduced Planck 
scale, $M_F = M_*$, which provides the cut off for the low-energy effective theory.  We will see that heavy particles needed to cancel anomalies 
associated with the gauged U(1)$_F$ flavor symmetry appear at two intermediate scales associated with the expectation values of the $s$ and $s'$ fields.

The gauge quantum numbers of the relevant fields are shown in the first two rows of Table~\ref{table:gaugec}.  Aside from the two scalars,
$s$ and $s'$, and the right-handed bottom quark, $d_R^3$, all other fields shown are heavy fermions that are chiral under U(1)$_F$ and
vector-like under the standard model gauge group; they become massive when the U(1)$_F$ symmetry is spontaneously broken.  It is straightforward to  check that all the gauge and gravitational anomalies are cancelled, with the parameters $N$ and $x$ unspecified.   
\begin{table}[t] 
\begin{center}
\begin{tabular}{| c |  c |  c |  c | c | c |}
\hline\hline
           & $s$  &   $s'$   & $d_R^3$  & $D_{R}^i$   &  $\overline{D_L^i}$    \\[7pt]  \hline
U(1)$_F$& $1$ &   $-\frac{1}{N}$ & $-1$         & $\frac{1}{N}+x$  &  $-x$    \\[7pt]  \hline
SU(3)$_C \times$SU(2)$_W \times$U(1)$_Y$ & $(1,1,0)$  & $(1,1,0)$  & $ (3,1,-1/3)$ & $ (3,1,-1/3)$ & 
$(\overline{3},1,+1/3)$ \\ \hline
U(1)$\times$U(1)$'$ & $(1,0)$  & $(0,1)$  & (-1,0) & $ (0,-1)$ & $ (0,0)$  \\
\hline\hline
\end{tabular}
\end{center}
\begin{center}
\begin{tabular}{| c |  c |  c |  c | c | c | c | c | c |}
\hline\hline
           & $E_R$ &  $\overline{E_L}$  &   $\overline{E'^i_L} $  &  $E'^i_R$  &  $N_R^j$ & $\overline{N_L^j}$ & $N'^k _R$ & $\overline{N'^k_L}$ \\[7pt] \hline
U(1)$_F$& $0$ &  $+1$  &   $-\frac{1}{N}-x$  &  $x$  &  $-\frac{1}{N}-x$ & $x$ & $1$ & $0$     \\[7pt] \hline
SU(3)$_C \times$SU(2)$_W \times$U(1)$_Y$ & $(1,1,-1)$  & $(1,1,+1)$  & $ (1,1,+1)$ & $ (1,1,-1)$ & $(1,1,0)$ & $(1,1,0)$& $(1,1,0)$ & $(1,1,0)$ \\  \hline
U(1)$\times$U(1)$'$ & $(0,0)$  & $(1,0)$  & (0,1) & $ (0,0)$ & $ (0,1)$ & $(0,0)$ & $(1,0)$ & $(0,0)$  \\
\hline\hline
\end{tabular}
\caption {Charge assignments under the gauged flavor symmetry, U(1)$_F$, the standard model gauge group, SU(3)$_C \times$SU(2)$_W \times$U(1)$_Y$,  and the accidental global U(1)$\times$U(1)$'$ symmetries discussed in the text.  Indices range from $i=1\ldots N$, $j=1\ldots 2N$ and $k=1\dots 2$.  Aside from $d_R^3$, all other standard model fields are U(1)$_F$ singlets. The parameters $N$ and $x$ are determined later by phenomenological constraints.} \label{table:gaugec}
\end{center}
\end{table}
Note that $x$ indicates a vectorial gauge rotation on the heavy fields $D$, $E'$ and $N'$, in addition to what is implied by the other charges shown.

Let $V_0(s,s')$ represent the scalar potential including only the renormalizable terms.   For $N>3$, $V_0$ is only a function of  $s^*s$ and 
$s'^* s'$, leading to an accidental U(1)$\times$U(1)$'$ global symmetry corresponding to separate phase rotations on the two flavon fields.  We will normalize the global charges to be $(1,0)$ and $(0,1)$ for the $s$ and $s'$ fields, respectively.  Using notation similar to Ref.~\cite{Fukuda:2017ylt}, we adopt the nonlinear representation
\begin{equation}
s = \frac{1}{\sqrt{2}} f_a e^{i \tilde{a}/f_a} \,\,\,\,\, \mbox{ and } \,\,\,\,\,  s' = \frac{1}{\sqrt{2}} f_b e^{i \tilde{b}/f_b} \,\,\, ,
\end{equation}
where $ \langle s \rangle = f_a / \sqrt{2}$, and $ \langle s' \rangle = f_b / \sqrt{2}$.  Since $V_0$ is independent of the phases of $s$ and $s'$,
$\tilde{a}$ and $\tilde{b}$ are absent from the potential.  When the U(1)$_F$ symmetry is gauged, however, one linear combination becomes the longitudinal component of the massive flavor gauge boson, while the remaining massless degree of freedom represents the goldstone boson of a residual U(1) global symmetry.  This linear combination becomes evident from studying the kinetic terms for $s$ and $s'$:
\begin{equation}
|D_\mu s |^2 + |D_\mu s' |^2 = \frac{1}{2} (\partial_\mu \widetilde{a})^2+\frac{1}{2} (\partial_\mu \widetilde{b})^2
-g_F A^\mu \partial_\mu (q f_a \widetilde{a}  +q' f_b \widetilde{b}) + \frac{g_F^2}{2} (q^2 f_a^2 +q'^2 f_b^2) A_\mu A^\mu \,\,\, ,
\end{equation}
where $g_F$ is the flavor gauge coupling, the gauge charges of the $s$ and $s'$ fields are $q$ and $q'$, respectively, with $q=+1$ and $q' = -1/N$ for the model defined in Table~\ref{table:gaugec}. We immediately identify the eaten linear combination
\begin{equation}
b = \frac{1}{\sqrt{q^2 f_a^2 + q'^2 f_b^2}}  (q f_a \widetilde{a}  + q' f_b \widetilde{b}) \,\,\, .
\end{equation}
The orthogonal linear combination is the physical massless degree of freedom, the flavored axion
\begin{equation}
a = \frac{1}{\sqrt{q^2 f_a^2 + q'^2 f_b^2}}  (q' f_b \widetilde{a}  - q f_a \widetilde{b}) \,\,\, ,
\label{eq:theaxion}
\end{equation}
or inverting
\begin{equation}
\left(\begin{array}{c} \widetilde{a} \\ \widetilde{b} \end{array}\right) = \frac{1}{\sqrt{q^2 f_a^2 + q'^2 f_b^2}}
\left(\begin{array}{cc}  q' f_b & q f_a \\  -q f_a & q' f_b \end{array}\right) \left(\begin{array}{c} a \\ b \end{array}\right) \,\,\,.
\label{eq:invert}
\end{equation}
Under a U(1)$_F$ gauge transformation, the exponentiated fields shift $\widetilde{a}/f_a \rightarrow \widetilde{a}/f_a + q \, \alpha$ and  $\widetilde{b}/f_b \rightarrow \widetilde{b}/f_b + q' \alpha$.  It is shown in Ref.~\cite{Fukuda:2017ylt} that a shift of the 
axion field $a/F$ by $2 \pi$ connects two gauge equivalent points in $\widetilde{a}$-$\widetilde{b}$ space provided that
\begin{equation}
F \equiv \frac{f_a f_b}{\sqrt{q^2 f_a^2 + q'^2 f_b^2}} \,\,\,.
\label{eq:bigF}
\end{equation}
We omit a repetition of that discussion here, but use the quantity $F$ in our discussion below.

We next consider constraints on the parameters $N$ and $x$.  The coupling of the $\widetilde{a}$ and
$\widetilde{b}$ fields to gluons and photons is determined by the U(1)$\times$U(1)$'$ color and electromagnetic 
anomalies, respectively,
\begin{equation}
{\cal L} = \frac{\alpha_s}{8 \pi} \left(2 N_a \frac{\widetilde{a}}{f_a} + 2 N_b  \frac{\widetilde{b}}{f_b} \right) G_{\mu\nu} \widetilde{G}^{\mu\nu}
+ \frac{\alpha_{em}}{8 \pi} \left(2 E_a \frac{\widetilde{a}}{f_a} + 2 E_b  \frac{\widetilde{b}}{f_b} \right) F_{\mu\nu} \widetilde{F}^{\mu\nu} \,\,\, ,
\label{eq:ggdual}
\end{equation} 
where $G_{\mu\nu}$ and $F_{\mu\nu}$ are the gluon and photon field strength tensors.  The color anomaly factors are given by
\begin{equation}
2 N_a = 1 \,\,\,\,\, \mbox{ and } \,\,\,\,\, 2 N_b = N \,\,\,,
\end{equation}
and the electromagnetic by
\begin{equation}
2 E_a = -4/3  \,\,\,\,\, \mbox{ and } \,\,\,\,\, 2 E_b = -4/3\,N \,\,\,.
\end{equation}
Using Eqs.~(\ref{eq:theaxion}) and (\ref{eq:bigF}),  we may rewrite Eq.~(\ref{eq:ggdual}) as
\begin{equation}
{\cal L} = -\frac{\alpha_s}{8 \pi} \frac{a}{f_{A}} G_{\mu\nu} \widetilde{G}^{\mu\nu} 
+ \frac{4}{3} \frac{\alpha_{em}}{8 \pi} \frac{a}{f_{A}} F_{\mu\nu} \widetilde{F}^{\mu\nu} \,\,\, ,
\end{equation}
where $f_{A} \equiv F/N$.  The quantity $f_A$ is what should be compared to bounds on the decay constant in conventional
axion models.  For example, the cosmological bound on the axion relic abundance  $f_A < 10^{12}$~GeV places a bound on
the combination of $f_a$ and $f_b$ that appears in Eq.~(\ref{eq:bigF}).   We identify the $s$ field with the flavon in the
model of Ref.~\cite{Carone:2019lfc}, where a global fit gave
\begin{equation}
f_a \approx 10^{-2} M_F  \,\,\, .
\end{equation}
We fix  $f_a$ to this value with $M_F=M_*$, so that $f_a \approx 2\times 10^{16}$~GeV; one then finds that $f_A < 10^{12}$~GeV implies, for 
example that $f_b <10^{13}$~GeV when $N=10$.  Note that for $f_b$ at this upper limit, we can compute the location of the Landau pole for hypercharge, which we expect to be drastically reduced by the multiplicity of heavy charged particles; we find this scale 
$\Lambda_{LP} \approx 3 \times 10^{18}$~GeV, which nonetheless remains above the cutoff of our effective theory.  We discuss this computation more explicitly below.  

We next turn to the issue of axion quality.  The accidental global symmetry of the potential is broken by terms that are not functions of $s^*s$ and $s'^*s'$ alone.   The lowest order U(1)$_F$ gauge-invariant term of this form is
\begin{equation}
{\cal L}_{bad}= \frac{\xi}{M_*^{N-3}}\, s\,  s'^N  + \mbox{h.c.}\,\,\, ,
\end{equation}
where $\xi$ is an order-one coupling that is generally complex.  This contributes both to the axion mass as well as to a linear term in the axion potential:
\begin{equation}
V(a) = - \frac{\mbox{Im }\xi}{\mbox{Re }\xi} f_A \Delta m^2  \, a + \frac{1}{2} (m_0^2 + \Delta m^2) \, a^2  \,\,\, ,
\end{equation}
where $m_0$ is the standard QCD contribution to the axion mass, and
\begin{equation}
\Delta m^2 = \frac{\mbox{Re }\xi}{2^{(N-1)/2}} \frac{f_a f_b^N}{f_A^2 M_*^{N-3}} \,\,\,.
\end{equation}
The linear term will shift the minimum of the axion potential away from the origin, reintroducing a non-vanishing value of the
$\overline{\theta}$ parameter,
\begin{equation}
\overline{\theta} = \langle a \rangle / f_A =  \frac{\mbox{Im }\xi}{\mbox{Re }\xi} \frac{\Delta m^2}{m_0^2 +  \Delta m^2} \,\,\, .
\end{equation}
Applying the phenomenological bound $\overline{\theta}<10^{-10}$~\cite{Tanabashi:2018oca}, and assuming that the real and imaginary parts of $\xi$ are of order unity, one concludes that $\Delta m^2 / m_0^2 <10^{-10}$.   Using the following estimate for the QCD contribution~\cite{Tanabashi:2018oca}
\begin{equation}
m_0 = 5.691\left(\frac{10^9 \mbox{ GeV}}{f_A}\right) 10^{-3}\mbox{ eV}  \,\,\, ,
\label{eq:m0}
\end{equation}
as well as our previous choice of $f_a= 10^{-2} M_*$, we find that this bound implies
\begin{equation}
f_b < \left[3.2387 \times 10^{-13} \mbox{ GeV}^4\right]^{1/N} (\sqrt{2})^{1-\frac{1}{N}} M_*^{1-\frac{4}{N}} \,\,\,.
\label{eq:quality}
\end{equation}
If we saturate this bound with $f_a$ fixed as previously noted, the mass scales of the heavy particles that carry standard model
charges are fixed, since these are determined via the U(1)$_F$-invariant Yukawa couplings
\begin{equation}
{\cal L}_{mass} = \lambda_D s' \overline{D^i}_L D^i_R + \lambda_E s \overline{E}_R E_L 
+ \lambda_E' s' \overline{E'^i}_R E'^i_L + \mbox{ h.c.}\,\,\,,
\label{eq:heavy}
\end{equation}
with the sum on $i=1\dots N$ implied.  These will contribute significantly to the running of hypercharge so we must check that the associated Landau pole remains above the cut off of our effective theory.  To do so, we evaluate the one-loop renormalization group equations between each threshold
\begin{align}
\alpha_Y^{-1}(m_b) & = \alpha_Y^{-1}(m_Z) + \frac{b_{SM}}{2 \pi} \ln\left(\frac{m_b}{m_Z}\right) \,\,\, , \\
\alpha_Y^{-1}(m_a) & = \alpha_Y^{-1}(m_b) + \frac{b_{SM}+\Delta b_b}{2 \pi} \ln\left(\frac{m_a}{m_b}\right) \,\,\, ,\\
\alpha_Y^{-1}(\Lambda_{LP}) & = \alpha_Y^{-1}(m_a) 
+ \frac{b_{SM}+\Delta b_b+\Delta b_a}{2 \pi} \ln\left(\frac{\Lambda_{LP}}{m_a}\right) \,\,\, ,
\end{align}
where we define the location of the Landau pole by $\alpha_Y^{-1}(\Lambda_{LP})=0$ using the standard model normalization of 
hypercharge\footnote{Of course, $\alpha_Y$ will become nonperturbative before this point.  However the difference between defining the 
Landau pole scale by some large perturbative value of the coupling versus $\alpha_Y^{-1}=0$ is not significant given the rapid increase in
the coupling around its blow-up point.}, and where the particle content of Table~\ref{table:gaugec} gives the beta functions
\begin{align}
& b_{SM} = -\frac{41}{6} \,\,\, ,\\
& \Delta b_{b} =-\frac{16}{9}\, N \,\,\, ,\\
& \Delta b_{a} =-\frac{4}{3} \,\,\, .
\end{align}
Taking the heavy particle thresholds to be $m_a \approx f_a$ and $m_b \approx f_b$ and 
$\alpha_Y^{-1}(m_Z)=98.43$, we find the Landau pole locations shown in Table~\ref{table:LP}.
\begin{table}[t] 
\begin{center}
\begin{tabular}{| c |  c | c | c |  c | c | c | c | c|}
\hline\hline
\hspace{0.25em} $N$ \hspace{0.25em} & \hspace{0.25em}$f_b$ (GeV)\hspace{0.25em} & \hspace{0.25em}$f_A$ (GeV) \hspace{0.25em}& \hspace{0.25em}$\Lambda_{LP}$ (GeV) & \hspace*{2em} 
& \hspace{0.25em}$N$ \hspace{0.25em}& \hspace{0.25em}$f_b$ (GeV) \hspace{0.25em}& \hspace{0.25em}$f_A$ (GeV) \hspace{0.25em}& \hspace{0.25em}$\Lambda_{LP}$ 
(GeV)\hspace*{0.25em}\\  \hline
$6$   & $1.4\times 10^4$ & $2.3\times 10^3$& $2.98 \times 10^{18}$ & & $11$ &  $4.4 \times 10^{10}$ & 
$4.0\times 10^9$ & $2.93 \times 10^{18}$\\
$7$   & $1.5\times 10^6$ & $2.2\times 10^5$ &$2.97 \times 10^{18}$ & & $12$ & $2.0 \times 10^{11}$ & $1.7\times 10^{10}$ & $2.92 \times 10^{18}$\\
$8$   & $5.3\times 10^7$ & $6.6\times10^6$ & $2.96 \times 10^{18}$ & & $13$ & $7.0 \times 10^{11}$ & $5.4\times10^{10}$ &
$2.92 \times 10^{18}$ \\
$9$   & $8.2\times 10^8$ &  $9.1 \times 10^7$ & $2.95 \times 10^{18}$ & & $14$ & $2.1 \times 10^{12}$ & 
$1.5\times10^{11}$ &  $2.91 \times 10^{18}$ \\
$10$ & $7.4\times 10^9$ & $7.4 \times 10^8$ & $2.94 \times 10^{18}$ & & $15$ & $5.4 \times 10^{12}$ & 
$3.6\times 10^{11}$ & $2.91 \times 10^{18}$\\
\hline\hline
\end{tabular}
\end{center}
\caption{Values of $f_b$ that saturate the bound on axion quality given in Eq.~(\ref{eq:quality}) as a function of $N$, with the associated axion decay constant, as well as the Landau pole scale for standard model hypercharge.} \label{table:LP}
\end{table}
We see that the Landau pole remains above the cut off of our effective theory, $M_*$, for a wide range in $N$; the value for this 
scale remains roughly constant, with the accelerated running caused by the greater particle multiplicity compensated by the 
heavier particle thresholds, which also increase with $N$, as given by the axion quality bound in Eq.~(\ref{eq:quality}).  We don't 
have similar worries for the U(1)$_H$ gauge coupling since its value at low energies is not fixed phenomenologically and can be 
taken small enough to keep its Landau pole safely above the cut off.

Bounds on the axion-photon coupling, defined by $g_{a\gamma \gamma} \equiv -\frac{\alpha_{em}}{2\pi}\frac{4}{3}\frac{1}{f_A}$, are 
summarized by~\cite{Arias-Aragon:2017eww}
\begin{equation}
\begin{array}{lcl}
|g_{a\gamma \gamma}|  \alt 7 \times 10^{-11} \mbox{ GeV}^{-1} & \hspace{1em} & \mbox{for } m_a \alt 10~\mbox{meV}  \\
|g_{a\gamma \gamma}|  \alt 10^{-10} \text{ GeV}^{-1} &  &\text{for } 10\ \text{meV} \lesssim m_a \lesssim 10 \ \text{eV}, \\
|g_{a\gamma \gamma}| \ll 10^{-12} \text{ GeV}^{-1} & & \text{for } 10\ \text{eV} \lesssim m_a \lesssim 0.1 \ \text{GeV}, \\
|g_{a\gamma \gamma}| \alt 10^{-3} \mbox{ GeV}^{-1} & & \text{for } 0.1\mbox{ GeV} \alt m_a \alt 1 \mbox{ TeV} \,\,\, .
\end{array} \label{eq:aphotb}
\end{equation}
Using the values of $f_A$ shown in Table~\ref{table:LP}, as well as the estimate for the axion mass in Eq.~(\ref{eq:m0}), these 
bounds eliminate $N \leq 8$, so that $N \geq 9$ is necessary for a viable model.  

Finally we consider the value of the parameter $x$.   This is not determined by any of the issues discussed thus far since its 
value does not contribute to the anomalies of any global symmetries (it parameterizes a vector rather than axial vector phase 
rotation) and does not affect any of the mass terms in Eq.~(\ref{eq:heavy}).  It does, however, determine the dimensions of operators 
that contribute to mass mixing between the heavy and light fermion fields.  For example, in the colored sector and for the 
choice $x=-2/N$, we can write the following mass mixing terms 
\begin{equation}
{\cal L}_{mix} = h_i \frac{s s'^2}{M_*^2} \, \overline{D^i}_L d_R^3 + g_i \frac{s'^*}{M_*} \overline{Q^3}_L H D_R + \mbox{ h.c.} \,\,\, ,
\label{eq:lagmix}
\end{equation}
which lead to small mixing between the heavy and light down-type quarks\footnote{Here, $Q^3_L$ is the third-generation standard model quark 
doublet.}.   Treating the interactions in Eq.~(\ref{eq:lagmix}) as perturbations, the second one provides a decay channel for the heavy $D$ fermion via $D \rightarrow d\, h^0$, where $h^0$ is the 
standard model Higgs boson. For the choice $N=10$, the results in Table~\ref{table:LP} tell us that 
$\langle s'^* \rangle / M_* \approx 2.6 \times 10^{-9}$, from which we can estimate the partial lifetime
\begin{equation}
\tau(D \rightarrow d h^0) \approx 10^{-15} \mbox{ sec.}
\end{equation}
Other decay channels involving U(1)$_F$ gauge boson exchange are also possible.  The general point is that the heavy fermions 
have at least one chirality with color and electroweak quantum numbers that match those of a standard model fermion, which makes it 
possible to construct operators that lead to the rapid decays of these states.   As a result we do not have to worry about direct search 
limits and cosmological consequences of heavy, long-lived charged particles.  If dark matter consists, in part, of light, neutral fermions, in addition 
to the flaxion component, we expect that a similar decays of the heavy to light neutral states can also be arranged.   We will not consider the issue 
of the stability of the heavy states further, since even in the case where they are exactly stable, it is possible that their abundance might be 
so low after re-heating~\cite{Duerr:2017amf} that there would be no negative consequences as  far as direct searches or cosmology is concerned.

\section{Extension with Leptons} \label{sec:leptons}

The model presented in our previous work, Ref.~\cite{Carone:2019lfc}, applied the flavor group discussed in Sec.~\ref{sec:quarks} to 
both the quarks and leptons.    A global fit to quark and lepton masses and CKM mixing angles demonstrated the viability of the 
model, with a flavor scale of $M_F=4 \times 10^{16}$~GeV, and running between the flavor scale and the $Z$ boson mass taken into 
account.   Operator coefficients were found via this fit to be consistent with the expectations of naive dimensional analysis.

In this section, we present a model that is a closer match to the one of Ref.~\cite{Carone:2019lfc} in that both quarks and leptons are 
subject to the $T' \times Z_3 \times$U(1)$_F$ flavor symmetry and $M_F$ is again fixed to $4 \times 10^{16}$~GeV, with $f_a = 10^{-2} M_F$ 
as suggested by the fit results.  In this way, all the numerical results of Ref.~\cite{Carone:2019lfc} are unchanged.  We will assume the most 
general set of $M_F$-suppressed higher-dimension operators, including those that could spoil the axion quality.   Despite the fact that the
ultraviolet (UV) cut off $M_F$ is smaller than $M_*$, the flavor-scale assumed in our quark-sector model, we will find that axion quality is sufficiently preserved.

With $f_a$ and $M_F$ fixed, there are two remaining free parameters, $f_b$ and $N$, which will be constrained by
\begin{itemize}
 \item[(a)] the axion quality bound that we have previously derived, which is now written as 
\begin{equation}
f_b < \left[3.2387 \times 10^{-13} \mbox{ GeV}^4\right]^{1/N} (\sqrt{2})^{1-\frac{1}{N}} M_F^{1-\frac{4}{N}}, 
\label{eq:quality2}
\end{equation}
\item[(b)] axion dark matter: If the PQ symmetry breaking happens before the inflationary phase, the axion can account for the DM relic density for decay constants on the order $f_A \sim 10^{11} \mbox{ to } 10^{13}\text{ GeV}$~\cite{Preskill:1982cy,Abbott:1982af,Dine:1982ah} without fine tuning of the misalignment angle. However other production mechanisms can also be implemented that allow for a lower axion decay constant, see for example Refs.~\cite{Co:2017mop,Co:2019jts,Co:2018mho,Co:2018phi,Harigaya:2019qnl,Co:2020dya}. Thus we only impose the upper bound $f_A \leq 10^{13}$ GeV.  It is also possible that dark matter
has multiple components so that the relic density need not be saturated by the axion's contribution.
\item[(c)] the requirement that the Landau pole of the hypercharge gauge coupling remain above our UV cutoff, the flavor scale $M_F$.  This constraint is relevant given the multiplicity of states with non-zero hypercharge in our extended heavy sector. \end{itemize}
Besides the above constraints, there are also constraints from the flavor-changing couplings of the axion. It was shown in Ref.~\cite{Carone:2019lfc} that the most stringent limit comes from the meson decay $K^+ \rightarrow \pi^+ a$  (see Eq.~$(3.19)$ in that reference). Since the most relevant limit concern quarks lets focus on that sector for now.  Derivatively coupled, flavor-changing axion couplings were obtained in Ref.~\cite{Carone:2019lfc} by applying the nonlinear field redefinition
\begin{equation}
d_R^3 \rightarrow e^{-ia/f_a}d_R^3\,\,\, ,
\end{equation}
where $a$ was the axion field in that model, and then rotating to the quark mass eigenstate basis.  In the scenario we consider 
here, however, the analogous redefinition will involve the $\tilde{a}$ field instead, which is not the axion field.  Re-expressing the
derivative interaction in terms of the linear combination of the $\tilde{a}$ and $\tilde{b}$ fields identified with the axion ({\em c.f.}, Eqs.~(\ref{eq:theaxion}) and (\ref{eq:invert})), then the bound on $f_a$ given in $(3.19)$ of Ref.~\cite{Carone:2019lfc} is modified to:
\begin{equation}
f_A < \frac{f_a^2}{6.3 \times 10^{10} \text{GeV}}.
\end{equation}
which is trivially satisfied for our choice $f_a = 4 \times 10^{14}$~GeV. Therefore we will not be concerned by the flavor-changing neutral current constraints henceforth. We will show how other relevant constraints can be satisfied below. 

\subsection{The Model}
The scalar fields and colored fermions charged under the gauged $U(1)_F$ of our quark-sector model remain unchanged while new color singlets are introduced to cancel gauge anomalies. The charge assignments of this model are presented in Table~\ref{table:model2}.
\begin{table}[H] 
\begin{center}
\begin{tabular}{| c |  c |  c |  c | c | c | c |}
\hline\hline
           & $s$  &   $s'$   & $d_R^3$ & $L^3$ & $D_{R}^i$   &  $\overline{D_L^i}$        \\[7pt]  \hline
U(1)$_F$& $1$ &   $-\frac{1}{N}$ & $-1$  & $1$       & $\frac{1}{N}+x$  &  $-x$    \\[7pt]  \hline
SU(3)$_C \times$SU(2)$_W \times$U(1)$_Y$ & $(\mathbf{1},\mathbf{1},0)$  & $(\mathbf{1},\mathbf{1},0)$  & $ (\mathbf{3},\mathbf{1},-1/3)$ & $(\mathbf{1},\mathbf{2},-\frac{1}{2})$ & $ (\mathbf{3},\mathbf{1},-1/3)$ & 
$(\overline{\mathbf{3}},\mathbf{1},+1/3)$ \\ \hline
U(1)$\times$U(1)$'$ & $(1,0)$  & $(0,1)$  & $(-1,0)$ & $(1,0)$ & $ (0,-1)$ & $ (0,0)$      \\
\hline\hline
\end{tabular}
\end{center}
\begin{center}
\begin{tabular}{| c |  c |  c |  c | c | c | c | c | c | }
\hline\hline
                                                                           &$\lambda^i_L$   &$\lambda^i_R $     &  $F_L^j$     &                                                                           $F_R^j$ &   $G_L^k$ & $G_R^k$  \\[7pt] \hline
U(1)$_F$                                   &$-x -\frac{1}{N}$                                     &$-x$                                                    &  $-1$  &  $0$ & $x+\frac{1}{N}$ & $x$     \\[7pt] \hline
SU(3)$_C \times$SU(2)$_W \times$U(1)$_Y$  &$(\mathbf{1},\mathbf{2},-\frac{1}{2})$& $(\mathbf{1},\mathbf{2},-\frac{1}{2})$& $(\mathbf{1},\mathbf{1},0)$ & $(\mathbf{1},\mathbf{1},0)$ & $(\mathbf{1},\mathbf{1},0)$& $(\mathbf{1},\mathbf{1},0)$  \\  \hline
U(1)$\times$U(1)$'$                                  &$(0,1)$          & (0,0) & $ (-1,0)$ & $ (0,0)$ & $(0,-1)$ & $(0,0)$   \\
\hline\hline
\end{tabular}
\caption {Charge assignments under the gauged flavor symmetry, U(1)$_F$, the standard model gauge group, 
SU(3)$_C \times$SU(2)$_W \times$U(1)$_Y$,  and the accidental global U(1)$\times$U(1)$'$ symmetries discussed in the text.  Indices 
range from $i=1\ldots N$, $j=1\ldots 5$ and $k=1\dots 5N$.  The fields $d_R^3$ and $L^3$ represent third-generation standard model 
fields; all other standard model fields are U(1)$_F$ singlets. The parameters $N$ and $x$ play the same role as in the quark-sector
model discussed in Sec.~\ref{sec:quarks}.} \label{table:model2}
\end{center}
\end{table}
In this model the heavy $\lambda_L$ and  $\lambda_R$ fields transform in the fundamental representation of $SU(2)_W$. The extra fermion 
exotics, $F_{L/R}$ and $G_{L/R}$,  cancel the $U(1)_F^3$ and $U(1)_F\times \text{Grav}^2$ gauge anomalies and are neutral under the SM 
gauge group. 

Mass terms for the exotics are given by 
\begin{equation}
\mathcal{L}_1 =  s' (  \kappa_1  \overline{D}_L D_R +\kappa_2  \overline{\lambda}_L \lambda_R +\kappa_3 \overline{G}_R G_L ) 
        +s  \ \kappa_4 \overline{F}_R F_L   + \text{h.c.},\label{Yukawas}
\end{equation}
where the $\kappa$'s are Yukawa couplings and the flavor indices on the heavy fields are omitted.  From this expression, we see how 
the accidental U(1) and U(1)$'$ symmetries of the scalar potential may be extended to the Yukawa couplings, with the global charges identified in 
the third row of Table~\ref{table:model2}.

The induced axion coupling to the $G\widetilde{G}$ term is given by the same formulas presented in the last section since the charges of 
the colored fermions under the accidental $U(1)\times U(1)'$ group are the same. However the axion coupling to photon pairs will be 
modified by the differences in the heavy particle spectra, including the presence of the new heavy leptons that are doublets under 
SU(2)$_W$ in the present theory.  For each of the U(1) global symmetries there is an $ F \widetilde{F}$ interaction corresponding to the associated anomaly.  These are given by~\cite{Carone:2019lfc} 
\begin{equation}
{\cal L} \supseteq \frac{\alpha_{em}}{8 \pi} \left[   \frac{\tilde{a}}{f_{a}} (2N_B + N_W)_{U(1)}    +     \frac{\tilde{b}}{f_{b}} (2N_B + N_W)_{U(1)'}   \right] F_{\mu\nu} \widetilde{F}^{\mu\nu} \,\,\, ,
\end{equation}
where $N_B$ and $N_W$ are the anomaly coefficients for hypercharge and isospin respectively. The value of these coefficients is completely determined once the charges of the scalar fields are fixed. Using the values presented in Table~\ref{table:model2} one obtains
\begin{equation}
(2N_B + N_W)_{U(1)} = \frac{8}{3},
\end{equation}
 \begin{equation}
(2N_B + N_W)_{U(1)'} = \frac{8}{3}N,
\end{equation}
leading to the axion-photon coupling
\begin{equation}
{\cal L} \supseteq - \frac{\alpha_{em}}{8 \pi}\frac{8}{3} \frac{a}{f_A} F_{\mu\nu} \widetilde{F}^{\mu\nu}.
\end{equation}
Note that the numerical coefficient is the same as what one would find in the simplest DFSZ axion models~\cite{DFSZ}.

\subsection{Model Constraints}
Since the exotic fermions with non-zero hypercharge, in this case $D^i$ and $\lambda^i$, for $i=1 \ldots N$, obtain their masses from the same scalar, the running of the hypercharge gauge coupling will be modified above the threshold given approximately by the scalar $s'$ vev.   This is different from the model introduced in the last section where the heavy particles with hypercharge appeared at two distinct energy thresholds.  At 1-loop order,  the location of the Landau pole is determined here by
\begin{align}
\alpha_Y^{-1}(m_b) & = \alpha_Y^{-1}(m_Z) + \frac{b_{SM}}{2 \pi} \ln\left(\frac{m_b}{m_Z}\right), \\
\alpha_Y^{-1}(\Lambda_{LP}) & = \alpha_Y^{-1}(m_b)
+ \frac{b_{SM}+\Delta b_b}{2 \pi} \ln\left(\frac{\Lambda_{LP}}{m_b}\right),
\end{align}
where the contribution to the beta function is 
\begin{equation}
\Delta b_b = - \frac{10}{9}N.
\end{equation}
Analogous to Table~\ref{table:LP}, we present the location of the Landau pole for different heavy particle multiplicities $N$, assuming
that the scale $f_b$ saturates the axion quality condition, Eq.~\eqref{eq:quality2}. We also show the predicted value of the axion decay constant $f_A$. 
\begin{table}[H] 
\begin{center}
\begin{tabular}{| c |  c |  c | c |  c | c | c | c | c|}
\hline\hline
\hspace{0.25em}$N$\hspace{0.25em}  & \hspace{0.25em}$f_b$ (GeV) \hspace{0.25em}& \hspace{0.25em}$f_A$ (GeV) \hspace{0.25em}& \hspace{0.25em}$\Lambda_{LP}$ (GeV)\hspace*{0.25em} & \hspace{2em} & \hspace{0.25em}$N$\hspace{0.25em} & \hspace{0.25em}$f_b$ (GeV) \hspace{0.25em}& \hspace{0.25em}$f_A$ (GeV) \hspace{0.25em}
& \hspace{0.25em}$\Lambda_{LP}$ (GeV)\hspace*{0.25em} \\  \hline
$6$   & $3.7\times 10^3$ &  $630 \ $&$4.5 \times 10^{22}$ & & $11$ &  $3.7 \times 10^{9}$ & $3.3\times 10^8$&$8.6 \times 10^{20}$\\
$7$   & $2.8\times 10^5$ &  $4.1 \times 10^{4}\ $&$1.6 \times 10^{22}$ & & $12$ & $1.5 \times 10^{10}$ & $1.2\times 10^9$&$5.1 \times 10^{20}$\\
$8$   & $7.4\times 10^6$ &  $9.3 \times 10^{5}\ $&$6.6 \times 10^{21}$ & & $13$ & $4.7 \times 10^{10}$ & $3.6\times 10^9$&$3.1 \times 10^{20}$\\
$9$   & $9.3\times 10^7$ &  $1.0 \times 10^{7}\ $&$3.0 \times 10^{21}$ & & $14$ & $1.3 \times 10^{11}$ & $9.1\times 10^9$&$2.0 \times 10^{20}$\\
$10$ & $7.0\times 10^8$ &  $7.0 \times 10^{7}\ $&$1.6 \times 10^{21}$ & & $15$ & $3.0 \times 10^{11}$ &$2.0\times 10^{10}$& $1.4 \times 10^{20}$\\
\hline\hline
\end{tabular}
\end{center}
\caption{Values of $f_b$ that saturate the bound on axion quality given in Eq.~(\ref{eq:quality2}) as a function of $N$, with
the associated value of the axion decay constant and the Landau pole scale for standard model hypercharge.} \label{table:LP2}
\end{table}
Table~\ref{table:LP2} shows that the Landau pole always remains above the UV cut off for the range in $N$ shown; in fact it is farther 
above the cut off than our earlier quark-sector model.   The bounds on the axion-photon coupling that were quoted in Eq.~(\ref{eq:aphotb}), 
apply here to the quantity $g_{a\gamma \gamma} \equiv \frac{\alpha_{em}}{2\pi}\frac{8}{3}\frac{1}{f_A}$.  Again, using the estimate for the 
axion mass in Eq.~(\ref{eq:m0}), one finds that the rows of Table~\ref{table:LP2} with  $N \leq 9$ are ruled out.  We thus find that $N \geq 10$ is 
necessary, similar to our quark sector-model.

\section{Discussion and Conclusions} \label{sec:conc}

It has been long argued that the Peccei-Quinn (PQ) solution to the strong CP problem could be spoiled by the presence of 
higher-dimension operators that violate the PQ symmetry explicitly unless their accompanying dimensionless coefficients are unnaturally 
small, or if the operators arise at sufficiently high order~\cite{Barr:1992qq}.   In this paper, we have extended the flavorful
axion model presented in our previous work~\cite{Carone:2019lfc} to address this problem by implementing a general strategy
for preserving axion quality proposed in Ref.~\cite{Fukuda:2017ylt}, the ``gauged Peccei-Quinn" mechanism.  The basic structure of our 
extended flavor sector was illustrated in a model of quark flavor presented in Sec.~\ref{sec:quarks}.   In that model, we made the 
theoretically economical choice of identifying the flavor scale $M_F$ (the cut off of our effective theory) with the reduced Planck scale 
$M_*$.   We then considered a more comprehensive model that included the charged leptons, with a flavor scale below the Planck scale;  
our choice of $M_F = 4 \times 10^{16}$~GeV as well as the various scales of flavor symmetry breaking were selected to match those of the 
model in Ref.~\cite{Carone:2019lfc}, so that the results of the global fit to fermion masses and mixing angles presented in that work would 
trivially carry over to the present  case.   However, since the gauged Peccei-Quinn mechanism renders the axion a linear combination of 
two scalar fields, some parametric differences in the flavor-changing axion couplings arise relative to our earlier results~\cite{Carone:2019lfc}; taking these into account, we showed that the most stringent bound from strange meson decays was trivially satisfied.   We also 
showed that the ratio of the electromagnetic to color anomalies in this model was the same as in the original model proposed in 
Ref.~\cite{Carone:2019lfc}, ({\em i.e.}, $8/3$), the same as the prediction of the simplest DFSZ axion models~\cite{DFSZ}.   We confirmed that both models we presented were consistent with the relevant low-energy constraints on the flaxion couplings for a range of model parameters. Since our model involved a relatively large sector of heavy fermions, some charged under U(1)$_Y$ and all vector-like under the standard model 
gauge group, we considered the accelerated running of the hypercharge gauge coupling at higher energy scales, confirming that its 
Landau pole remains above the ultraviolet cut off of our effective theory.    It is likely that flavor models exist that allow a simpler
adaptation of the mechanism of Ref.~\cite{Fukuda:2017ylt} to address the problem of flaxion quality.  Finding the models that allow
the simplest implementation may provide a clue as to which flaxion models are more likely to be realized in nature.


\begin{acknowledgments}  
We thank the NSF for support under Grant PHY-1819575.    
\end{acknowledgments}


\end{document}